\documentclass[twoside]{article}
\usepackage{amsmath}
\catcode`\@=11
\long\def\@makefntext#1{
\protect\noindent \hbox to 3.2pt {\hskip-.9pt
$^{{\eightrm\@thefnmark}}$\hfil}#1\hfill}       

\def\@makefnmark{\hbox to 0pt{$^{\@thefnmark}$\hss}}    

\def\ps@myheadings{\let\@mkboth\@gobbletwo
\def\@oddhead{\hbox{}
\rightmark\hfil\eightrm\thepage}
\def\@oddfoot{}\def\@evenhead{\eightrm\thepage\hfil
\leftmark\hbox{}}\def\@evenfoot{}
\def\sectionmark##1{}\def\subsectionmark##1{}}

\oddsidemargin=\evensidemargin
\newcounter{sectionc}\newcounter{subsectionc}\newcounter{subsubsectionc}
\renewcommand{\section}[1] {\vspace{12pt}\addtocounter{sectionc}{1}
\setcounter{subsectionc}{0}\setcounter{subsubsectionc}{0}\noindent
    {\tenbf\thesectionc. #1}\par\vspace{5pt}}
\renewcommand{\subsection}[1] {\vspace{12pt}\addtocounter{subsectionc}{1}
\setcounter{subsubsectionc}{0}\noindent
{\bf\thesectionc.\thesubsectionc. {\kern1pt \bfit #1}}\par\vspace{5pt}}
\renewcommand{\subsubsection}[1] {\vspace{12pt}\addtocounter{subsubsectionc}{1}
    \noindent{\tenrm\thesectionc.\thesubsectionc.\thesubsubsectionc.
    {\kern1pt \tenit #1}}\par\vspace{5pt}}

\newcounter{appendixc}
\newcounter{subappendixc}[appendixc]
\newcounter{subsubappendixc}[subappendixc]
\renewcommand{\thesubappendixc}{\Alph{appendixc}.\arabic{subappendixc}}
\renewcommand{\thesubsubappendixc}
    {\Alph{appendixc}.\arabic{subappendixc}.\arabic{subsubappendixc}}

\renewcommand{\appendix}[1] {\vspace{12pt}
        \refstepcounter{appendixc}
        \setcounter{figure}{0}
        \setcounter{table}{0}
        \setcounter{lemma}{0}
        \setcounter{theorem}{0}
        \setcounter{corollary}{0}
        \setcounter{definition}{0}
        \setcounter{equation}{0}
        \renewcommand{\thefigure}{\Alph{appendixc}.\arabic{figure}}
        \renewcommand{\thetable}{\Alph{appendixc}.\arabic{table}}
        \renewcommand{\theappendixc}{\Alph{appendixc}}
        \renewcommand{\thelemma}{\Alph{appendixc}.\arabic{lemma}}
        \renewcommand{\thetheorem}{\Alph{appendixc}.\arabic{theorem}}
        \renewcommand{\thedefinition}{\Alph{appendixc}.\arabic{definition}}
        \renewcommand{\thecorollary}{\Alph{appendixc}.\arabic{corollary}}
        \renewcommand{\theequation}{\Alph{appendixc}.\arabic{equation}}
        \noindent{\tenbf Appendix \theappendixc #1}\par\vspace{5pt}}
\newcommand{\subappendix}[1] {\vspace{12pt}
        \refstepcounter{subappendixc}
        \noindent{\bf Appendix \thesubappendixc. {\kern1pt \bfit #1}}
    \par\vspace{5pt}}
\newcommand{\subsubappendix}[1] {\vspace{12pt}
        \refstepcounter{subsubappendixc}
        \noindent{\rm Appendix \thesubsubappendixc. {\kern1pt \tenit #1}}
    \par\vspace{5pt}}

\newcommand{\textlineskip}{\baselineskip=13pt}
\newcommand{\smalllineskip}{\baselineskip=10pt}

\newcommand{\copyrightheading}[1]
    {\vspace*{-2.5cm}\smalllineskip{\flushleft
    {\footnotesize {\sl Quantum Information and Computation, Vol.~3, No.~3 (2003) 193-202 #1}}\\
    {\footnotesize \copyright\kern2pt Rinton Press}\\
     }}

\def\abstracts#1#2#3{{
    \centering{\begin{minipage}{4.5in}\footnotesize\baselineskip=10pt
    \parindent=0pt #1\par
    \parindent=15pt #2\par
    \parindent=15pt #3
    \end{minipage}}\par}}

\def\keywords#1{{
    \centering{\begin{minipage}{4.5in}\footnotesize\baselineskip=10pt
    {\footnotesize\it Keywords}\/: #1
     \end{minipage}}\par}}

\renewenvironment{thebibliography}[1]
        {\frenchspacing
     \ninerm\baselineskip=11pt
         \begin{list}{\arabic{enumi}.}
        {\usecounter{enumi}\setlength{\parsep}{0pt}
     \setlength{\leftmargin 12.7pt}{\rightmargin 0pt}
         \setlength{\itemsep}{0pt} \settowidth
    {\labelwidth}{#1.}\sloppy}}{\end{list}}

\newcounter{itemlistc}
\newcounter{romanlistc}
\newcounter{alphlistc}
\newcounter{arabiclistc}

\newcommand{\fcaption}[1]{
        \refstepcounter{figure}
        \setbox\@tempboxa = \hbox{\footnotesize Fig.~\thefigure. #1}
        \ifdim \wd\@tempboxa > 5in
           {\begin{center}
        \parbox{5in}{\footnotesize\smalllineskip Fig.~\thefigure. #1}
            \end{center}}
        \else
             {\begin{center}
             {\footnotesize Fig.~\thefigure. #1}
              \end{center}}
        \fi}

\newcommand{\tcaption}[1]{
        \refstepcounter{table}
        \setbox\@tempboxa = \hbox{\footnotesize Table~\thetable. #1}
        \ifdim \wd\@tempboxa > 5in
           {\begin{center}
        \parbox{5in}{\footnotesize\smalllineskip Table~\thetable. #1}
            \end{center}}
        \else
             {\begin{center}
             {\footnotesize Table~\thetable. #1}
              \end{center}}
        \fi}

\def\pmb#1{\setbox0=\hbox{#1}
    \kern-.025em\copy0\kern-\wd0
    \kern.05em\copy0\kern-\wd0
    \kern-.025em\raise.0433em\box0}

\def\fnt#1#2{\footnotetext{\kern-.3em
    {$^{\mbox{\scriptsize #1}}$}{#2}}}

\def\fpage#1{\begingroup
\voffset=.3in
\thispagestyle{empty}\begin{table}[b]\centerline{\footnotesize #1}
    \end{table}\endgroup}

\def\runninghead#1#2{\pagestyle{myheadings}
\markboth{{\protect\footnotesize\it{\quad #1}}\hfill}
{\hfill{\protect\footnotesize\it{#2\quad}}}}
\font\tenrm=cmr10
\font\tenit=cmti10
\font\tenbf=cmbx10
\font\bfit=cmbxti10 at 10pt
\font\ninerm=cmr9

\font\eightrm=cmr8

\def\FigName{figure}%
\newbox\captionbox
\long\def\@makecaption#1#2{%
  \ifx\FigName\@captype
    \vskip\abovecaptionskip
    \setbox\tempbox\hbox{{\figurecaptionfont #1\hskip1em #2}}
    \ifdim\wd\tempbox< 28pc
    \centerline{\box\tempbox}
    \else
    {\figurecaptionfont #1\hskip1em #2\par}
\fi\else
    \setbox\tempbox\hbox{{\tablecaptionfont #1\hskip1em #2}}
    \ifdim\wd\tempbox< 28pc
    \centerline{\box\tempbox}
    \else
    {\tablecaptionfont #1\hskip1em #2\par}%
    \fi
 \vskip\belowcaptionskip
 \fi}
\InputIfFileExists{psfig.sty}
{\typeout{^^Jpsfig.sty inputed...ok}}{\typeout{^^JWarning: psfig.sty could be be found.^^J}}
\InputIfFileExists{epsfsafe.tex}
{\typeout{^^Jepsfsafe.tex inputed...ok}}
            {\typeout{^^JWarning: epsfsafe.tex could not be found.^^J}}
\InputIfFileExists{epsfig.sty}
{\typeout{^^Jepsfig.sty inputed...ok}}{\typeout{^^JWarning: epsfig.sty could not be found.^^J}}
\InputIfFileExists{epsf.sty}
{\typeout{^^Jepsf.sty inputed...ok}}{\typeout{^^JWarning: epsf.sty could not be found.^^J}}%
\def\fps@figure{tbp}
\def\ftype@figure{1}
\def\ext@figure{lof}
\def\fnum@figure{Fig.\ \thefigure}
\def\qed{\hbox{${\vcenter{\vbox{              
   \hrule height 0.4pt\hbox{\vrule width 0.4pt height 6pt
   \kern5pt\vrule width 0.4pt}\hrule height 0.4pt}}}$}}

\begin{document}
\setlength{\textheight}{8.0truein}    

\runninghead{A matrix realignment method for recognizing entanglement}
            {Kai Chen and Ling-An Wu}

\normalsize\textlineskip
\thispagestyle{empty}
\setcounter{page}{1}


\fpage{1}
\centerline{\bf A MATRIX REALIGNMENT METHOD}
\vspace*{0.035truein}
\centerline{\bf FOR RECOGNIZING ENTANGLEMENT}
\vspace*{0.37truein}
\centerline{\footnotesize KAI CHEN\footnote{E-mail address:
kchen@aphy.iphy.ac.cn} and LING-AN WU\footnote{E-mail address: wula@aphy.iphy.ac.cn}}
\vspace*{0.015truein}
\centerline{\footnotesize\it Laboratory of Optical Physics, Institute of Physics}
\baselineskip=10pt
\centerline{\footnotesize\it Chinese Academy of Sciences, Beijing 100080, P.R. China}
\vspace*{0.225truein}

\vspace*{0.21truein}
\abstracts{ Motivated by the Kronecker
product approximation technique, we have developed a
very simple method to assess the inseparability of bipartite
quantum systems, which is based on a realigned matrix
constructed from the density matrix. For any separable
state, the sum of the singular values of the matrix should
be less than or equal to $1$. This condition provides a very
simple, computable necessary criterion for separability,
and shows powerful ability to identify most bound
entangled states discussed in the literature. As a byproduct of the
criterion, we give an estimate for the degree of entanglement
of the quantum state.}{}{}

\vspace*{10pt}
\vspace*{3pt}

\vspace*{1pt}\textlineskip  
\section{Introduction}           
\label{sec1}
\vspace*{-0.5pt}
Quantum entangled states have recently emerged as basic resources in the
rapidly expanding field of quantum information processing, with remarkable
applications such as quantum teleportation, cryptography, dense coding and
parallel computation  \cite{pre98,nielsen,zeilinger}. However, two fundamental questions
have only been partially answered: how do we know if a given
quantum state is entangled, and how entangled is it
still after interacting with a noisy environment?

From a practical point of view, the state of a composite quantum system is
called \emph{unentangled} or \emph{separable} if it can be prepared in a ``
\emph{local}'' or ``\emph{classical}'' way. A separable bipartite system can
be expressed as an ensemble realization of pure product states $\left| \psi
_{i}\right\rangle _{A}\left| \phi _{i}\right\rangle _{B}$ occurring with a
certain probability $p_{i}$:
\begin{equation}
\rho _{AB}=\sum_{i}p_{i}\rho _{i}^{A}\otimes \rho _{i}^{B},  \label{sep}
\end{equation}
where $\rho _{i}^{A}=\left| \psi _{i}\right\rangle _{A}\left\langle \psi
_{i}\right| $, $\rho _{i}^{B}=\left| \phi _{i}\right\rangle _{B}\left\langle
\phi _{i}\right| $, $\sum_{i}p_{i}=1$, and $\left| \psi _{i}\right\rangle
_{A}$, $\left| \phi _{i}\right\rangle _{B}$ are normalized pure states of
subsystems $A$ and $B$, respectively \cite{werner89}. If no convex linear
combination exists for a given $\rho _{AB}$, then the state is called
``\emph{entangled}''.

There have been considerable efforts in recent years to
analyze the separability and quantitative character of
quantum entanglement. For a pure state $\rho _{AB}$, it is
separable iff $\rho _{AB}\equiv \rho _{A}\otimes \rho _{B}$
where $\rho _{A,B}$ is the reduced density matrix defined as
$\rho _{A}=Tr_{B}\rho _{AB}$ and $\rho _{B}=Tr_{A}\rho
_{AB}$. However, for a generic mixed state $\rho _{AB}$,
finding a decomposition as in Eq.~(\ref{sep}) or proving
that it does not exist is a non-trivial task (see
\cite{3hreview} and references therein). The first important
breakthrough was made by Peres who proposed that partial
transposition with respect to one subsystem of the density
matrix for a separable state is positive, i.e., has
non-negative eigenvalues \cite{peres}. Now known as the $PPT$
criterion, this was shown by Horodecki \textit{et al}
to be sufficient for bipartite systems of $2\times 2$
and $2\times 3$ \cite{3hPLA223}. In the same paper,
a necessary and sufficient condition for
separability was found by establishing a close connection between
positive map theory and separability. Soon
after, Wootters succeeded in computing the
``\textit{entanglement of formation}" \cite{be96} and thus
obtained a separability criterion for $2\times 2$ mixtures
\cite{wo98}. The ``\textit{reduction criterion}'' proposed
independently in \cite{2hPRA99} and \cite{cag99} gives
another necessary criterion which is equivalent to the $PPT$
criterion for $2\times n$ composite systems but is generally
weaker than the $PPT$ criterion. Pittenger \textit{et al}
gave also a sufficient criterion for separability connected
with the Fourier representations of density matrices
\cite{Rubin00}. Later, Nielsen \textit{et al}
\cite{nielson01} presented another necessary criterion
called the \textit{majorization criterion}: the decreasingly
ordered vector of the eigenvalues for $\rho_{AB}$ is
majorized by that of $\rho_{A}$ or $\rho_{B}$ alone for a
separable state. A new method of constructing
\textit{entanglement witnesses} for detecting entanglement
was given in \cite{3hPLA223} and \cite{ter00,lkch00}. There
are also some necessary and sufficient criteria of
separability for low rank cases of the density matrix, as
shown in \cite{hlvc00,afg01}. In addition, it was shown in
\cite{wu00} and \cite{pxchen01} that a necessary and
sufficient separability criterion is also equivalent to
certain sets of equations.

However, despite these advances, a practical computable criterion
for generic bipartite systems is mainly limited to the $PPT$,
reduction and majorization criteria, as well as a recent extension
of the $PPT$ criterion based on semidefinite programs \cite{dps}.
The $PPT$ criterion has been considered a strong one up to now, but
in general it is still not sufficient for higher dimensions of
greater than or equivalent to $3$. Several counterexamples of
\textit{bound entangled states} with $PPT$ properties were provided
in \cite{hPLA97}, and their entanglement does not seem to be
``useful'' for distillation \cite{benPRL96}. These states are
``weakly'' inseparable and it is very hard to establish with
certainty their inseparability \cite{cag99}.

In this paper we focus on an inseparability test for a generic
bipartite quantum system of arbitrary dimensions. The basic procedure is
the same as that of the cross norm criterion presented in \cite{ru02}
but is mathematically much more straightforward. Motivated by
the Kronecker product approximation of the density matrix
\cite{loan,pits} we have derived a
directly computational method to recognize entangled states based
on a realigned matrix constructed from the density matrix. An
estimate is also given for depicting the ``\textit{distance}'' from
a quantum state to the maximally entangled state and to the
maximally mixed state. Several typical examples of bound entangled
states are shown to be recognized by this criterion in Section 3. A
brief summary and discussion are given in the last section.

\section{\label{sec2} A necessary separability criterion
based on a matrix realignment method}
In this section we will present an inseparability criterion to
recognize entangled states based on simple matrix analysis.
Some of its characteristics are shown in several propositions. At
the same time, we give a possible measure for the degree of
entanglement as a by-product. The main tools used are a
technique developed by Loan and Pitsianis \cite{loan,pits}
for the Kronecker product approximation of a given matrix,
and some results from matrix analysis (see Chapters $3$ and
$4$ of \cite{hornt} for a more extensive background). We
shall first review some of the notation and results required
in this paper.

\vspace*{12pt}
\noindent
\emph{\textbf{Definition:}} \emph{For each $m\times n$ matrix $A=[a_{ij}]$,
where $a_{ij}$ is the matrix entry of A, we define the vector $vec(A)$ as}
\begin{equation*}
vec(A)=[a_{11},\cdots ,a_{m1},a_{12},\cdots ,a_{m2},\cdots ,a_{1n},\cdots
,a_{mn}]^{T}.
\end{equation*}

\vspace*{12pt}
\noindent
Let $Z$ be an $m\times m$ block matrix with block size $n\times n$. We
define a realigned matrix $\widetilde{Z}$ of size $m^{2}\times n^{2}$ that
contains the same elements as $Z$ but in different positions as
\begin{equation}
\widetilde{Z}\equiv\left[
\begin{array}[c]{c}
vec(Z_{1,1})^{T}\\
\vdots\\
vec(Z_{m,1})^{T}\\
\vdots\\
vec(Z_{1,m})^{T}\\
\vdots\\
vec(Z_{m,m})^{T}
\end{array}
\right], \label{trans}
\end{equation}
so that the singular value decomposition for $\widetilde{Z}$ is
\begin{equation}
\widetilde{Z}=U\Sigma V^{\dagger }=\sum_{i=1}^{q}\sigma
_{i}u_{i}v_{i}^{\dagger },  \label{singular}
\end{equation}
where $U=[u_{1}u_{2}\cdots u_{m^{2}}]\in {\mathcal{C}}^{m^{2}\times m^{2}}$
and $V=[v_{1}v_{2}\cdots v_{n^{2}}]\in {\mathcal{C}}^{n^{2}\times n^{2}}$
are unitary, $\Sigma $ is a diagonal matrix with elements $\sigma _{1}\geq
\sigma _{2}\geq \cdots \geq \sigma _{q}\geq 0$ and $q=\min (m^{2},n^{2})$.
In fact, the number of nonzero singular values $\sigma _{i}$ is the rank $r$
of matrix $\widetilde{Z}$, and $\sigma _{i}$ are exactly the nonnegative
square roots of the eigenvalues of $\widetilde{Z}\widetilde{Z}^{\dagger }$
or $\widetilde{Z}^{\dagger }\widetilde{Z}$ \cite{hornt}. Based on the above
constructions, Loan and Pitsianis obtained the following representation for
$Z$
\begin{equation}
Z=\sum_{i=1}^{r}(X_{i}\otimes Y_{i}),  \label{sum}
\end{equation}
with $vec(X_{i})=\sqrt{\sigma _{i}}u_{i}$ and $vec(Y_{i})=\sqrt{\sigma _{i}}
v_{i}^{\ast }$ \cite{loan,pits}.

For any given density matrix $\rho _{AB}$ we can associate a realigned
version $\widetilde{\rho _{AB}}$ according to the transformation of
Eq.~(\ref{trans}). For example, a $2\times2$ bipartite
density matrix $\rho$ can be transformed as:
\begin{align}
\rho &  =\left(
\begin{array}
[c]{cc|cc}
\rho_{11} & \rho_{12} & \rho_{13} & \rho_{14}\\
\rho_{21} & \rho_{22} & \rho_{23} & \rho_{24}\\\cline{1-4}
\rho_{31} & \rho_{32} & \rho_{33} & \rho_{34}\\
\rho_{41} & \rho_{42} & \rho_{43} & \rho_{44}
\end{array}
\right)   \longrightarrow\widetilde{\rho}=\left(
\begin{array}
[c]{cccc}
\rho_{11} & \rho_{21} & \rho_{12} & \rho_{22}\\\cline{1-4}
\rho_{31} & \rho_{41} & \rho_{32} & \rho_{42}\\\cline{1-4}
\rho_{13} & \rho_{23} & \rho_{14} & \rho_{24}\\\cline{1-4}
\rho_{33} & \rho_{43} & \rho_{34} & \rho_{44}
\end{array}
\right)  .
\end{align}

For any separable system $\rho _{AB}$, we arrive at the following
main theorem:

\vspace*{12pt}
\noindent
{\bf Theorem:} \label{theorem} \emph{If an $m\times n$
bipartite density matrix $\rho _{AB}$ is separable, then for the
$m^{2}\times n^{2}$ matrix $\widetilde{\rho _{AB}}$ the \textsl{Ky Fan} norm
$N(\widetilde{\rho _{AB}})\equiv \sum_{i=1}^{q}\sigma _{i}(\widetilde{\rho
_{AB}}),$ which is the sum of all the singular values of $\widetilde{\rho
_{AB}}$, is $\leq 1,$ or equivalently $\log (N(\widetilde{\rho _{AB}}
))\leq 0$, where $q=\min (m^{2},n^{2})$.}

\vspace*{12pt}
\noindent
{\bf Proof:} Suppose $\rho _{AB}$ has a decomposition of $\rho
_{AB}=\sum_{i}p_{i}\rho _{i}^{A}\otimes \rho
_{i}^{B}=\sum_{i}p_{i}(U_{i}^{A}E_{11}^{(m,m)}U_{i}^{A}{}^{\dagger })\otimes
(V_{i}^{B}E_{11}^{(n,n)}V_{i}^{B}{}^{\dagger })$ with $0\leq p_{i}\leq 1$
satisfying $\sum_{i}p_{i}=1$. Here $E_{ij}^{(k,l)}$ is a $k\times l$ matrix
which has entry $1$ in position $i,j$ and all other entries as zero. The
$U_{i}^{A,B}$ are the unitary matrices which diagonalize $\rho _{i}^{A,B}$.
Applying the properties of Kronecker products:
\begin{eqnarray*}
vec(XYZ) &=&(Z^{T}\otimes X)vec(Y),\text{ \ \ \ \ \ \ \ \ (see \cite{hornt})}
\\
Z=X\otimes Y &\Longleftrightarrow &\widetilde{Z}=vec(X)vec(Y)^{T},\text{ \ \
\ (see \cite{pits})}
\end{eqnarray*}
we have
\begin{eqnarray}
&&\widetilde{\rho _{i}^{A}\otimes \rho _{i}^{B}}=vec(\rho _{i}^{A})vec(\rho
_{i}^{B})^{T}  \notag \\
&=&(U_{i}^{A}{}^{\ast }\otimes
U_{i}^{A})vec(E_{11}^{(m,m)})((U_{i}^{B}{}^{\ast }\otimes
U_{i}^{B})vec(E_{11}^{(n,n)}))^{T}  \notag \\
&=&(U_{i}^{A}{}^{\ast }\otimes
U_{i}^{A})vec(E_{11}^{(m,m)})vec(E_{11}^{(n,n)})^{T}(U_{i}^{B}{}^{\dagger
}\otimes U_{i}^{B}{}^{T}).  \notag \\
&&  \label{unitary}
\end{eqnarray}
Since $U_{i}^{A}$ and $U_{i}^{B}$ are unitary, it is evident that
$U_{i}^{A}{}^{\ast }\otimes U_{i}^{A}$ and $U_{i}^{B}{}^{\dagger }\otimes
U_{i}^{B}{}^{T}$ are unitary also. Moreover,
$vec(E_{11}^{(m,m)})vec(E_{11}^{(n,n)})^{T}=E_{11}^{(m^{2},n^{2})}$ has a
unique singular value $1$. Then $\widetilde{\rho _{i}^{A}\otimes \rho
_{i}^{B}}$ has also a unique singular value $1$ due to the fact that the
\textsl{Ky Fan} norm is unitarily invariant \cite{hornt}. Therefore, as a
convex linear combination of $\widetilde{\rho _{i}^{A}\otimes \rho _{i}^{B}}$
, $\widetilde{\rho _{AB}}$ should have the \textsl{Ky Fan} norm
$N(\widetilde{\rho _{AB}})\leq \sum_{i}p_{i}N(\widetilde{\rho _{i}^{A}\otimes
\rho _{i}^{B}})=\sum_{i}p_{i}=1$. Here we have used the inequality
$\sum_{i}^{q}\sigma _{i}(A+B)\leq \sum_{i}^{q}\sigma
_{i}(A)+\sum_{i}^{q}\sigma _{i}(B)$ where $A$ and $B$ are $k\times l$
matrices and $q=\min (k,l)$ \cite{hornt}. \hfill \qed\,

The cross norm criterion proposed in \cite{ru02} leads to the
same result as our above Theorem, but with the expressions in Dirac
bra-ket notation. Here we obtain identical results but from simple matrix
analysis. Moreover, our approach has special advantages in
its concise and explicit expressions: the two factor spaces $A$ and
$B$ just correspond to different rows and columns of
$\widetilde{\rho _{AB}}$ under the transformation $\rho
_{AB}\longrightarrow \widetilde{\rho _{AB}}$. Also the technology
for the proof is much simpler compared with the complicated
operator algebra used in \cite{ru02}. Now we
only need to rearrange the entries of $\rho _{AB}$ according to
Eq.~(\ref{trans}), then compare the sum of the square roots of the
eigenvalues for $\widetilde{\rho _{AB}}\widetilde{\rho
_{AB}}^{\dagger }$ (i.e., the \emph{Ky Fan} norm or trace norm of
$\widetilde{\rho _{AB}}$) with $1$. The separability criterion is
strong enough to be sufficient for many lower dimensional systems.
It is straightforward to verify sufficiency for the Bell diagonal
states \cite{be96}, Werner states in dimension $d=2$
\cite{VoPRA01} and isotropic states in arbitrary dimensions
\cite{2hPRA99}, as was done in \cite{ru02}. So far we have not yet
found a close connection between this criterion and the $PPT$
criterion, but from Eq.~(\ref{trans}) we have the compatibility
relation $N(\widetilde{\rho })=N( \widetilde{\rho ^{T_{A}}})$,
since $\widetilde{\rho }$ and $\widetilde{\rho ^{T_{A}}}$ only
differ up to a series of elementary row transformations which are
naturally unitary and keep the \textsl{Ky Fan} norm invariant.
For pure states, we have the following stronger conclusion:

\vspace*{12pt}
\noindent
{\bf Proposition~1:} \emph{A
bipartite pure state is separable iff $\widetilde{\rho _{AB}}$
has a unique singular value 1.}

\vspace*{12pt}
\noindent
{\bf Proof:} \textsl{Necessity:} Given a bipartite separable
pure state we have $\rho_{AB}=\rho ^{A}\otimes \rho ^{B}$.
From Eq.~(\ref{unitary}), $\widetilde{\rho _{AB}}$ has a
unique singular value 1.

\noindent
\textsl{Sufficiency:}
$\widetilde{\rho _{AB}}$ has a unique singular value $\sigma _{1}=1,$ so
$\widetilde{\rho _{AB}}=\sigma _{1}u_{1}v_{1}^{\dagger }=u_{1}v_{1}^{\dagger
} $. By Eq.~(\ref{sum}), we have $\rho _{AB}=\alpha \otimes \beta $ with
$vec(\alpha )=u_{1}$ and $vec(\beta )=v_{1}$. Moreover, $\rho _{AB}=\rho
_{AB}^{2}\Longrightarrow \alpha \otimes \beta =\alpha ^{2}\otimes \beta ^{2}$
. Thus the eigenvalues of $\alpha $ and $\beta $ should both be $1$ or $-1$
at the same time because $\rho _{AB}$ has only one eigenvalue $1$. This
yields $\rho ^{A}=\alpha $ and $\rho ^{B}=\beta $ if $\alpha $ and $\beta $
have eigenvalue $1$, and $\rho ^{A}=-\alpha $ and $\rho ^{B}=-\beta $ if
$\alpha $ and $\beta $ have eigenvalue $-1$. Therefore $\rho _{AB}$ is a pure
separable density matrix. \hfill \qed\,

\vspace*{12pt}
Now we derive a dual criterion based on the Theorem:

\vspace*{12pt}
\noindent
{\bf Corollary:} \label{corollary} \emph{For a
separable $m\times n$ system $\rho _{AB},$ a permuted version
$\rho _{BA}$ which exchanges the first and second  factor spaces of
$\rho _{AB}$ can be defined as
\begin{equation}
\rho _{BA}=S(n,m)\rho _{AB}S(m,n),  \label{ab2ba}
\end{equation}
where $S(m,n)=\sum_{i=1}^{m}\sum_{j=1}^{n}E_{ij}^{(m,n)}\otimes
(E_{ij}^{(m,n)})^{T}$. Then for the $n^{2}\times m^{2}$ realigned matrix
$\widetilde{\rho _{BA}}$ the \textsl{Ky Fan} norm $N(\widetilde{\rho _{BA}})$
is $\leq 1$ or equivalently $\log (N(\widetilde{\rho _{BA}}))\leq 0$.}

\vspace*{12pt}
\noindent
{\bf Proof:} For a separable $m\times n$ system $\rho _{AB}$, there exists
a decomposition: $\rho _{AB}=\sum_{i}p_{i}\rho _{i}^{A}\otimes \rho _{i}^{B}$.
Using the property $Y\otimes X=$ $S(n,m)(X\otimes Y)S(m,n)$ where $X$ is
an $m\times m$ matrix and $Y$ is an $n\times n$ one \cite{hornt}, we have
$\rho _{BA}=S(n,m)(\sum_{i}p_{i}\rho _{i}^{A}\otimes \rho
_{i}^{B})S(m,n)=\sum_{i}p_{i}\rho _{i}^{B}\otimes \rho _{i}^{A}$. This
density matrix is also separable but associated with an $n\times m$ system,
so we have $N(\widetilde{\rho _{BA}})\leq 1$ according to the Theorem.
\hfill \qed\,

\vspace*{12pt}
This criterion is equivalent to the Theorem, due to the property
$\widetilde{Y\otimes X}=vec(Y)vec(X)^{T}=(vec(X)vec(Y)^{T})^T=(\widetilde{X\otimes
Y})^T$. Thus we have $\widetilde{\rho _{AB}}=(\widetilde{\rho
_{AB}})^T$ and further $N(\widetilde{\rho
_{AB}})=N(\widetilde{\rho _{BA}})$. This can also be seen from the
symmetry with respect to the two subsystems of the cross norm
given in \cite{ru02}.

We expect that $\log N(\widetilde{\rho _{AB}})$ may depict a possible
measure of entanglement for the corresponding system. In fact, for
the $d-$dimension maximally mixed state
$\rho _{m}=Id/d^{2}$ it is easy to derive
$\widetilde{\rho_{m}}=\frac{1}{d}\sum_{i,j=0}^{d-1}|ii\rangle \langle jj|$
and $\log N(\widetilde{\rho _{m}})=-\log d$, while for the maximally entangled state
$\rho _{e}=\frac{1}{d}\sum_{i,j=0}^{d-1}|ii\rangle \langle jj|$ we have
$\widetilde{\rho _{e}}=Id/d^{2}$ and $\log N(\widetilde{\rho _{e}}
)=\log d$. Noticing the relations $\rho _{m}=\widetilde{\rho _{e}}$ and
$\widetilde{\rho _{m}}=\rho _{e}$, we see that they are dual and
$\log N(\widetilde{\rho })$ is distributed symmetrically about $0$ where the
separable pure states are located. Hence any bipartite quantum state $\rho $
can be depicted by such an approximate measure $\log N(\widetilde{\rho })$
representing its $``distance"$ from $\rho _{m}$ and $\rho _{e}$. Furthermore
we have the following property for $\log N(\widetilde{\rho })$:

\vspace*{12pt}
\noindent
{\bf Proposition~2:} \emph{Applying a local unitary transformation
leaves $\log N(\widetilde{\rho })$ invariant, i.e.
\begin{equation}
\log N(\widetilde{\rho _{AB}^{^{\prime }}})=\log N(\widetilde{\rho _{AB}}),
\end{equation}
where $\rho _{AB}^{^{\prime }}=(U\otimes V)\rho _{AB}(U^{\dagger }\otimes
V^{\dagger })$ and $U,V$ are unitary operators acting on the A,B subsystems,
respectively.}

\vspace*{12pt}
\noindent
{\bf Proof:} From Eq.~(\ref{sum}), we have a decomposition of $\rho
_{AB}=\sum_{i=1}^{q}\alpha _{i}\otimes \beta _{i}$ where $q=\min
(m^{2},n^{2})$. Setting $\gamma _{i}=(U\otimes V)(\alpha _{i}\otimes
\beta _{i})(U^{\dagger }\otimes V^{\dagger })$, we have
\begin{eqnarray}
\widetilde{\gamma _{i}} &=&vec(U\alpha _{i}U^{\dagger })vec(V\beta
_{i}V^{\dagger })^{T}  \notag \\
&=&(U^{\ast }\otimes U)(vec(\alpha _{i})vec(\beta _{i}))(V^{\dagger
}\otimes V^{T}).
\end{eqnarray}
Summing  all the components $\widetilde{\gamma _{i}}$, we have $\widetilde{\rho
_{AB}^{^{\prime }}}=(U^{\ast }\otimes U)\widetilde{\rho _{AB}}
(V^{\dagger }\otimes V^{T})$. Since $U$ and $V$ are unitary, it is evident
that $U^{\ast }\otimes U$ and $V^{\dagger }\otimes V^{T}$ are unitary,
hence $\log N(\widetilde{\rho
_{AB}^{^{\prime }}})=\log N(\widetilde{\rho _{AB}})$ since $N(\widetilde{\rho })$ is a unitary invariant
norm. \hfill \qed\,

\vspace*{12pt}\noindent
In addition, like any unitary invariant norm, $N(\widetilde{\rho
})$ is convex \cite{hornt}, but $\log N(\widetilde{\rho _{AB}})$
is not  convex due to the concave logarithmic function.
Therefore, this rough ``\emph{measure}'' is not exact because the
criterion is not a necessary and sufficient
one, but it depicts the entanglement to some degree in a subtle
way independent of the $PPT$ criterion.

\section{\label{sec3} Application of the criterion to
bound entangled states}
In this section we give some typical examples to display the power
of our separability criterion to distinguish the bound
entangled states from separable states.

\vspace*{12pt}
\noindent \emph{Example 1:} $3\times 3$ bound entangled states constructed
from \emph{unextendible product bases} (UPB)

In \cite{UPB}, Bennett \textit{et al} introduced a $3\times 3$
inseparable bound entangled state from the following bases:
\begin{eqnarray*}
{|\psi _{0}\rangle } &=&{\frac{1}{\sqrt{2}}}{|0\rangle }({|0\rangle }-{\
|1\rangle }),\ \ {|\psi _{1}\rangle }={\frac{1}{\sqrt{2}}}({|0\rangle }-{\
|1\rangle }){|2\rangle }, \\
{|\psi _{2}\rangle } &=&{\frac{1}{\sqrt{2}}|2\rangle }({|1\rangle }-{\
|2\rangle }),\text{ \ }{|\psi _{3}\rangle }={\frac{1}{\sqrt{2}}}({|1\rangle }
-{|2\rangle }){|0\rangle }, \\
{|\psi _{4}\rangle } &=&{\frac{1}{3}}({|0\rangle }+{|1\rangle }+{|2\rangle )}
({|0\rangle }+{|1\rangle }+{|2\rangle ),}
\end{eqnarray*}
from which the density matrix can be expressed as
\begin{equation}
\rho =\frac{1}{4}(Id-\sum_{i=0}^{4}{|\psi _{i}\rangle \langle \psi
_{i}|}).  \label{upb1}
\end{equation}
Direct computation gives $\log N(\widetilde{\rho })\doteq 0.121$, which
shows that there is slight entanglement in this state.

When the UPB have the following construction \cite{UPB}
\begin{equation*}
|\psi _{j}\rangle =|\vec{v}_{j}\rangle \otimes |\vec{v}_{2j \bmod
5} \rangle ,\;\;j=0,\ldots ,4
\end{equation*}
where the vectors $\vec{v}_{j}$ are
$\vec{v}_{j}=N(\cos {{\frac{2\pi j}{5}}},\sin {\frac{{2\pi j}}{5}},h)$,
with $j=0,\ldots ,4$ and $h={\frac{1}{2}}\sqrt{1+\sqrt{5}}$ and
$N=2/\sqrt{5+\sqrt{5}}$, then the $PPT$ state of Eq.~(\ref{upb1})
gives $\log N(\widetilde{\rho })\doteq 0.134$, which
identifies this bound entangled state.

\vspace*{12pt}
\noindent \emph{Example 2:} Horodecki $3\times 3$ bound entangled state

\begin{figure} [htbp]
\vspace*{13pt} \centerline{\psfig{file=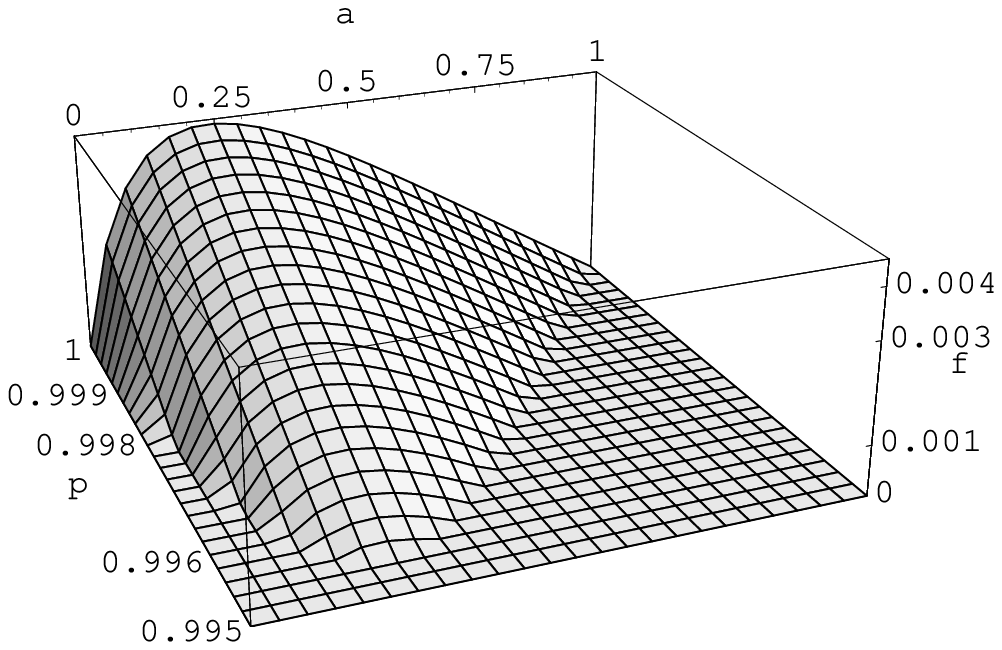, width=8.2cm}}
\vspace*{13pt}
\fcaption{\label{fig1} Estimate
of degree of entanglement for a Horodecki $3\times 3$ bound
entangled state. When $a=0.236$ we have maximal entanglement
of the state for $0.9955 <p \leq 1$.}
\end{figure}

Horodecki gives a very interesting weakly inseparable state in \cite{hPLA97}
which cannot be detected by the $PPT$ criterion. The density matrix $\rho $
is real and symmetric:
\begin{equation}
\rho ={\frac{1}{8a+1}}\left[
\begin{array}{ccccccccc}
a & 0 & 0 & 0 & a & 0 & 0 & 0 & a \\
0 & a & 0 & 0 & 0 & 0 & 0 & 0 & 0 \\
0 & 0 & a & 0 & 0 & 0 & 0 & 0 & 0 \\
0 & 0 & 0 & a & 0 & 0 & 0 & 0 & 0 \\
a & 0 & 0 & 0 & a & 0 & 0 & 0 & a \\
0 & 0 & 0 & 0 & 0 & a & 0 & 0 & 0 \\
0 & 0 & 0 & 0 & 0 & 0 & {\frac{1+a}{2}} & 0 & {\frac{\sqrt{1-a^{2}}}{2}} \\
0 & 0 & 0 & 0 & 0 & 0 & 0 & a & 0 \\
a & 0 & 0 & 0 & a & 0 & {\frac{\sqrt{1-a^{2}}}{2}} & 0 & {\frac{1+a}{2}}
\end{array}
\right], \label{bes2}
\end{equation}
where $0<a<1$. To compare our result with the computation for the
\emph{entanglement of formation} in \cite{moorPRA01}, we consider a mixture of the
Horodecki state and the maximally mixed state
$\rho _{p}=p\rho +(1-p)Id/9$, and show the function
\begin{equation}
f=\max (0,\log N(\widetilde{\rho _{p}}))
\end{equation}
in Fig.\ref{fig1}.

\noindent We see that $\log $$N(\widetilde{\rho _{p}})>0$
when $p=1$ and $0<a<1$, which completely identifies these
bound entangled states. Furthermore, when $a=0.236$, $\log
N(\widetilde{\rho _{p}})$ has a maximum of $0.0044$ and
follows a similar trend to that of the degree of
entanglement for $\rho _{p}$ (with a maximum at $a=0.225$)
given in \cite{moorPRA01}. It can be seen that we obtain an
upper bound $p=0.9955$ for $ \rho _{p}$ which still has
entanglement when $a=0.236$.

To test the criterion, we programmed a routine to analyze
various $PPT$ entangled states described in the literature. On a
$600$MHz desktop computer we performed a systematic search
by checking $10^{5}$ randomly chosen examples of the seven
parameter family of $PPT$ entangled states in~\cite{bruss}
within half an hour. The Theorem could detect entanglement in
about $21\%$ of these bound entangled states satisfying $\rho=\rho ^{T_A}$
and gave a maximum value for $\log N(\widetilde{\rho })$ of $0.22$, which
is a convincing demonstration of the strength of the criterion.

\vspace{0.6cm}
{\bf \noindent Remark 1:}
\begin{quotation}
One might expect the Theorem to be equivalent to the
$PPT$ criterion for a $2\times 2$ system. Unfortunately, this is
not the case, and certain $2\times2$ entangled states cannot
be recognized by our criterion. We
take the following example from \cite{3hPLA223}
which is a separable state only when $p=\frac{1}{2}$ or $a=0$ or
$a=1$:
\begin{figure} [htbp]
\vspace*{13pt}
\centerline{\psfig{file=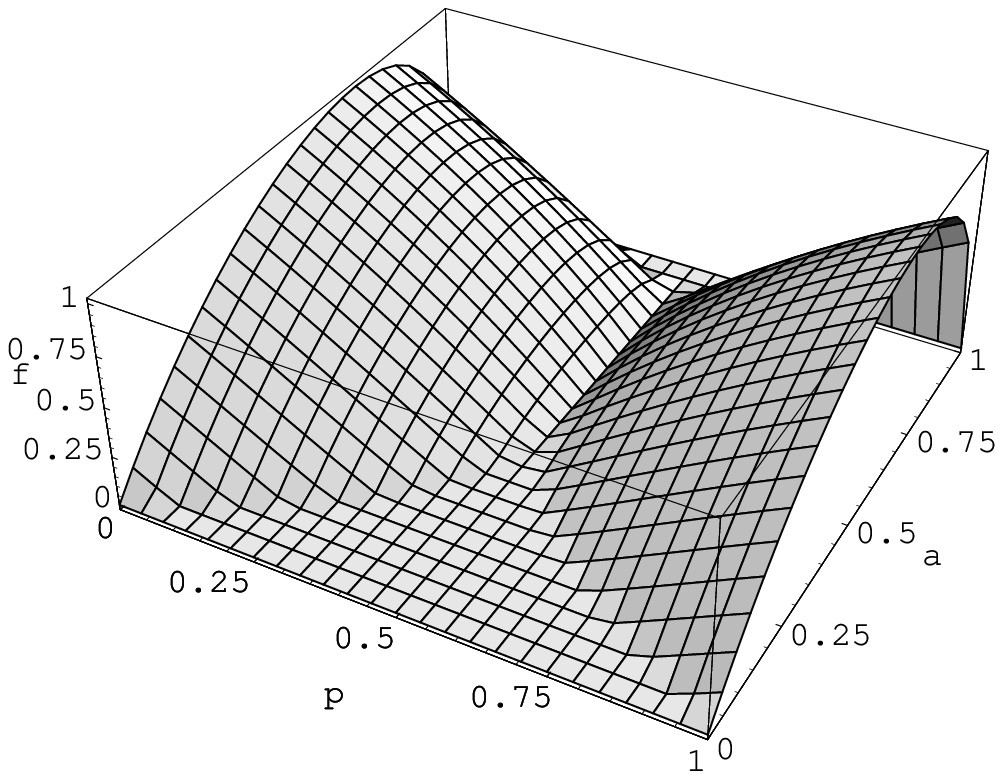, width=8.2cm}}
\vspace*{13pt}
\fcaption{\label{fig2} The function
$f=\max (0,\log N(\widetilde{\rho}))$ for the quantum state
of Eq.~\ref{22state}.}
\end{figure}
\begin{eqnarray}
\rho=\left(
\begin{array}[c]{cccc}
pa^{2} & 0 & 0 & pab\\
0 & (1-p)a^{2} & (1-p)ab & 0\\
0 & (1-p)ab & (1-p)b^{2} & 0\\
pab & 0 & 0 & pb^{2}
\end{array}
\right)
\label{22state}
\end{eqnarray}
where $a,b>0$ and $\left| a\right| ^{2}+\left| b\right|
^{2}=1$. We plot the value of $f=\max (0,\log
N(\widetilde{\rho}))$ in Fig.~\ref{fig2} with respect to $a$
and $p$. We can see that, apart from when $p=\frac{1}{2}$
or $a=0$ or $1$, there are still some regions where
$f=0$. Thus these $2\times2$ entangled states cannot be completely
detected by our criterion.
\end{quotation}
\vspace{0.6cm}
{\bf \noindent Remark 2:}
\begin{quotation}
It is interesting to make a comparison between
$\log N(\widetilde{\rho})$ or $N(\widetilde{\rho})-1$
with the entanglement of formation $E_f$ \cite{be96}. By direct comparison, for the density matrix
of Eq.~\ref{bes2}, $\log N(\widetilde{\rho})$ or $N(\widetilde{\rho})-1$
is less than the entanglement of formation calculated in \cite{moorPRA01}. As for the state of a $2\times 2$ system,
we can conveniently calculate the entanglement of formation
by Wootter's formula \cite{wo98}.
By straightforward computation, we have
\begin{enumerate}
  \item For $2\times 2$ entangled Werner states \cite{VoPRA01}:
$\log N(\widetilde{\rho})\geq E_f=N(\widetilde{\rho})-1$;
  \item For the density matrix of Eq.~\ref{22state}:
$N(\widetilde{\rho})-1\leq E_f$  but there is no definite relation between
$\log N(\widetilde{\rho})$ and $E_f$;
  \item For $3\times 3$ entangled isotropic states \cite{2hPRA99}:
$N(\widetilde{\rho})-1\geq E_f$ and $\log N(\widetilde{\rho})\geq E_f$.
\end{enumerate}
Thus we find that there is no ordered relationship between $E_f$ and
$N(\widetilde{\rho})-1$ or $\log N(\widetilde{\rho})$.
\end{quotation}

\section{\label{sec4} Summary and discussion}

Summarizing, we have presented a matrix realignment method to
assess the separability of the density matrix for a bipartite
quantum system in arbitrary dimensions. The criterion provides a
necessary condition for separability and is quite easy to compute.
It shows remarkable ability to recognize most known bound entangled
states in the literature where the $PPT$ test fails. Moreover, it
gives a rough estimate for the degree of entanglement. We also show
that this criterion is not equivalent to $PPT$ even for $2\times 2$
quantum states. Moreover, there is no definite relationship between our
estimate and the entanglement of formation. However, in combination
with the $PPT$ criterion, we can significantly expand our ability
to distinguish directly the entanglement and separability of any
quantum state in arbitrary dimensions. Our method has recently been
developed further by a linear contraction approach that gives a
generic separability criterion \cite{Horo02}. It is expected that
this method will find more applications in the study of
multipartite quantum systems and other problems in quantum
information theory and quantum computation. We leave the explicit
physical meaning of the criterion as an open question that awaits
further study.

\vspace*{12pt}
\noindent{\bf Acknowledgements}

\vspace*{12pt}
K.C. would like to thank Dr. L. Yang for valuable
discussions and Prof. Guozhen Yang for encouragement. This
work was supported by the Chinese Academy of Sciences, the
National Program for Fundamental Research, the National
Natural Science Foundation of China Grants 19974073 and
60178013 and the China Postdoctoral Science Foundation. The
authors also greatly appreciate the anonymous referees'
valuable suggestions for improving the original version of
this paper.

\end{document}